\documentstyle[12pt]{article}
\date{}
\begin{document}

\title{ Comparison of the $\pi$ and $\Delta$ operator approximations \\
for the two-dimensional $t$-$J$ model}
\author{Youngho Park\\
Institute of Physics, Academia Sinica, Nankang, Taipei 11529, Taiwan}
\maketitle
\begin{abstract}
We compare the spectra of the new $\pi$ operator of the SO($5$) theory
and the conventional $\Delta$ operator for the two-dimensional $t$-$J$ model.
We also calculate the weight transferred to the two-hole ground state from
half-filling by these operators.
We find that the spectra of these operators are quite similar and
the weight for the $\pi$ operator is smaller than the weight for the $\Delta$
operator.  We argue that the two-dimensional $t$-$J$ model does not
have a good approximate SO($5$) symmetry claimed in Ref. \cite{Eder}.
\end{abstract}


\newpage

The low-energy states of the two-dimensional $t$-$J$ model
have been often related to the antiferromagnetism and the high-$T_c$ 
superconductivity of the cuprate material \cite{Dagotto}.
The SO($5$) theory relates the spin excited state of the antiferromagnetic
half-filling and the hole-doped $d$-wave ground state of the 
high-$T_c$ material by the $\pi$ operator \cite{Zhang}.
In other words, according to the SO($5$) theory in microscopic model
Hamiltonians such as the two-dimensional $t$-$J$ and Hubbard models these two
different states can be related by an SO($5$) symmetry group.
Since the neutron scattering resonance in the cuprate YBCO was interpreted
as the $\pi$ resonance, a low-energy resonance associated with the
$\pi$ operator \cite{Delmer}, the basic idea of the SO($5$) theory has
been highly controversial \cite{GBA,ZL}.
Some numerical studies have been done to support this theory by finding the
$\pi$ resonance for the dynamical correlation functions of the
$\pi^\dagger$ ($\pi$) operator
for the $t$-$J$ and Hubbard models \cite{Eder,Meixner,Hanke}.
Eder {\it et al.} claimed that low-energy states of the $t$-$J$ model
form SO($5$) symmetry multiplets and the hole-doped ground states away from 
half-filling are obtained from the higher-spin states at half-filling through
SO($5$) rotations \cite{Eder}.
For example, to get the two-hole ground state we apply the $\pi$ operator to
the $S=1$ state at half-filling.  Therefore,
one may claim that this $\pi$ operator approximation is an alternative of the
conventional $d$-wave pairing operator, $\Delta$ operator approximation for the
hole-doped ground state.
However, in order to check how good the new one, the $\pi$ operator
approximation is, it is necessary to compare with the old one, the $\Delta$
operator approximation.  
We will make qualitative comparison, that is,  comparison of
the spectra of these two operators and quantitative comparison, that is, 
comparison of the weight transferred to the two-hole ground state from
half-filling by these operators.
We will also discuss about the approximate SO($5$) symmetry of the
two-dimensional $t$-$J$ model claimed by Eder
{\it et al.} \cite{Eder}.

The the $t$-$J$ model Hamiltonian is
\begin{equation}
H=-t\sum_{\langle i,j\rangle,\sigma}
(\tilde{c}_{i\sigma}^\dagger \tilde{c}_{j\sigma} + {\rm H.c.})
+J\sum_{\langle i,j\rangle}({\bf S}_i\cdot {\bf S}_j-\frac{1}{4} n_in_j),
\end{equation}
where the sum is over nearest-neighbor sites and $\tilde{c}_{i\sigma}=
c_{i\sigma}(1-n_{i\sigma})$ is the electron annihilation operator with
the constraint of no double occupancy.
The $\pi$ and $\Delta$ operators are expressed as follows \cite{Eder,Zhang}:
\begin{equation}
\pi_\alpha=\sum_{\bf p} (\cos p_x-\cos p_y)
c_{{\bf p}+{\bf Q}i}(\sigma^\alpha\sigma^y)_{ij} c_{-{\bf p}j},
\end{equation}
\begin{equation}
\Delta=\sum_{\bf p} (\cos p_x-\cos p_y)
c_{{\bf p}\uparrow}c_{-{\bf p}\downarrow},
\end{equation}
where $\sigma^\alpha$ is the vector of Pauli matrices and
 ${\bf Q}=(\pi,\pi)$.
The $\pi$ operator carries charge $-2$, a spin triplet and momentum
transfer $\Delta {\bf P}=(\pi,\pi)$, and has $d$-wave symmetry.
On the other hand, the $\Delta$ operator is the same but it carries a spin
singlet and $\Delta {\bf P}=(0,0)$.
While the $\Delta$ operator relates the the half-filled ground state which 
is a spin singlet and the two-hole ground state which is also a spin singlet,
the $\pi$ operator does not relates these two states directly instead,
one has to go through the spin excited state.
In order to get the two-hole ground state ($|\psi_{2h,S=0}\rangle$),
we apply the $\pi$ operator to the minimum state with the total spin
$S=1$ and the total momentum ${\bf P}=(\pi,\pi)$ at half-filling
($|\psi_{{\rm HF},S=1}\rangle$),
or we apply the $\Delta$ operator to the half-filled ground state
($|\psi_{{\rm HF},S=0}\rangle$) \cite{Poilblanc}.
The former case belongs to the SO($5$) allowed transition since both
$|\psi_{{\rm HF},S=1}\rangle$ and $|\psi_{2h,S=0}\rangle$ are members of
the same $\nu$ irrep SO(5) multiplet according to Eder {\it et al.} 
\cite{Eder}.

The spectra of these operators can be calculated as follows:
\begin{equation}
A_{\hat{\pi}} = -\frac{1}{\pi} {\rm Im}\langle\psi_{{\rm HF},S=1}|
\hat{\pi}^\dagger
\frac{1}{\omega-H-E_{2h}+i\epsilon}\hat{\pi}|\psi_{{\rm HF},S=1}\rangle,
\end{equation}
\begin{equation}
A_{\hat{\Delta}} = -\frac{1}{\pi} {\rm Im}\langle\psi_{{\rm HF},S=0}|
\hat{\Delta}^\dagger
\frac{1}{\omega-H-E_{2h}+i\epsilon}\hat{\Delta}|\psi_{{\rm HF},S=0}\rangle,
\end{equation}
where $E_{2h}$ is the two-hole ground state energy.
We show the spectra for the $18$-site lattice in Fig. $1$.
Here, we choose $\pi_\pm$ operator and $S_z=\pm 1$ state, which would give
favorable result in the spin-polarized sector than using $\pi_z$ operator and
$S_z=0$ state.
The spectra of the $\pi$ and $\Delta$ operators are quite similar.
The reason is that the final states are in the same sector,
$N_h=-2$ (two holes), ${\bf P}=(0,0)$, and both operators
have $d$-wave symmetry.  For the $18$-site lattice, both
$|\psi_{{\rm HF},S=1}\rangle$ and $|\psi_{{\rm HF},S=0}\rangle$ are
$s$-wave and $|\psi_{2h,S=0}\rangle$ is $d$-wave.
They have a single dominant low-energy peak and high-energy incoherent part.
The low-energy peak corresponds to the two-hole ground state.
The intensity of the low-energy peak increases as increasing $J/t$ and
is slightly bigger for the spectra of the $\Delta$ operator than for the
spectra of the $\pi$ operator.  This is the case for all values of $J/t$.
However, when we work with $\pi_z$ operator and $S_z=0$ state instead of
$\pi_\pm$ operator and $S_z=\pm 1$ state
we find the intensity of the low-energy peak is even lower, which can be
seen from the calculations of the transferred weight later.
Since the $\Delta$ operator does not change $S$ of the state, the
high-energy incoherent part in the spectra of the $\Delta$ operator will
include excited states with $S=0$.  
On the other hand, the spectra of the $\pi$ operator will include
higher $S$ states which are also transferred from
$|\psi_{{\rm HF},S=1}\rangle$ by the $\pi$ operator.
For example, we identify the third peak for $J=0.5$ and $J=0.75$ cases as a
$S=2$, $d$-wave state which is not found in the spectra of the $\Delta$
operator.

With knowing the fact that there is a low-energy resonance for the $\Delta$
operator at half-filling \cite{Dagotto,Poilblanc}, if one choose an
appropriate operator, the $\pi$ operator and a starting state with relevant
quantum numbers and symmetry, $|\psi_{{\rm HF},S=1}\rangle$ in order to
reach to the same final sector as having the $\Delta$ operator and
$|\psi_{{\rm HF},S=0}\rangle$, it is not a surprise to observe a low-energy
resonance at the same energy as for the $\Delta$ operator case.
The important thing is how big the resonance is, that is, how well we
can approximate the two-hole ground state by these operators starting
from different states.
Similarly, one can do the same thing for the final sector, $S=1$ and
${\bf P}=(\pi,\pi)$ at half-filling with the spin density wave operator
$S_{\bf Q}^{+}=\sum_{\bf{p}}
c_{{\bf p}+{\bf Q}\uparrow}^\dagger c_{{\bf p}\downarrow}$
and $|\psi_{{\rm HF},S=0}\rangle$ , or with the $\pi^\dagger$ operator and
$|\psi_{2h,S=0}\rangle$ as in Ref. \cite{Eder} as well as for the
Hubbard model \cite{Meixner}.
The low-energy resonance for the $S_{\bf Q}^{+}$ operator and
$|\psi_{{\rm HF},S=0}\rangle$  is apparently at
the energy of a magnon excitation, $\Delta E \sim J$ (or $\sim 4t^2/U$ for
the Hubbard model) relative to the half-filled ground state energy.
So is for the $\pi^\dagger$ operator and $|\psi_{2h,S=0}\rangle$, hence the
$\pi$ resonance.

The intensity of the low-energy peak is proportional to
the weight transferred to the two-hole ground state from half-filling
by the operators.
We calculate the transferred weight for the $\pi$ operator, for both $\pi_z$
and $\pi_\pm$, and for the $\Delta$ operator.
That is, we calculate the following quantities \cite{Poilblanc}:
\begin{equation}
W_\pi=\frac{|\langle\psi_{2h,S=0}|\pi|\psi_{{\rm HF},S=1}\rangle|}
{\sqrt{\langle\psi_{{\rm HF},S=1}|\pi^+\pi|\psi_{{\rm HF},S=1}\rangle}},
\end{equation}
\begin{equation}
W_\Delta=\frac{|\langle\psi_{2h,S=0}|\Delta|\psi_{{\rm HF},S=0}\rangle|}
{\sqrt{\langle\psi_{{\rm HF},S=0}|\Delta^+\Delta|\psi_{{\rm HF},S=0}\rangle}}.
\end{equation}
In Fig. $2$, we plot $W_{\pi_\pm}$, $W_{\pi_z}$ and $W_\Delta$  for the
$18$-site lattice.  All weights increase as increasing $J/t$.
The square values of these are almost linear in $J/t$ for this range of
$J/t$ \cite{Poilblanc}.
Both $W_{\pi_\pm}$ and $W_{\pi_z}$ are smaller than $W_\Delta$ for all values
of $J/t$ but $W_{\pi_\pm}$ is closer to $W_\Delta$ than $W_{\pi_z}$.
This result can be understood by the fact that unlike the $\Delta$ operator
case some weight is transferred to higher $S$ states by the $\pi$ operator
though higher $S$ states, especially states with the appropriate $d$-wave
symmetry, are energetically much higher and do not seem to have too much
weight.
We also calculate the weights for the $20$-site lattice and show in 
Fig. $3$.
The result remains qualitatively the same.
These results indicate that the approximation by the $\pi$ operator is not
better than the approximation by the $\Delta$ operator for the two-hole
ground state regardless of the lattice size.  
Also, one may expect similar results for larger hole-doping.
In order to get the ground state at larger hole-doping from half-filling
one must apply the $\pi$ operator successively and moreover, one has to go
through the spin-excited state each time.
On the other hand, one can apply the $\Delta$ operator successively to achieve
the same thing.  Therefore, one can easily derive that the $\pi$ operator
approximation can not be better than the $\Delta$ operator approximation for
any hole-doping.

Finally, we would like to discuss about the approximate SO($5$) symmetry for
the two-dimensional $t$-$J$ model claimed by Eder {\it et al.} \cite{Eder}.
Concerning their claim that ``low-energy states of the $t$-$J$ model 
form SO($5$) symmetry multiplets'', it should be noted that the SO($5$)
multiplets include only $s$-wave and $d$-wave states since the $\pi$ operator
connects only states which differ by $d$-wave symmetry with each other.
However,
there are other symmetry states such as $p$-wave state which is abundant in the
low-energy part \cite{res} and seemingly becomes the ground state for small
$J$ in the thermodynamic limit \cite{Hamer}.
Then the idea is that ``at a chemical potential comparable to the mean level
spacing, the superspin multiplets are nearly degenerate''
and ``the variance of the
splitting among various states connected by the $\pi$ operator is a
well-defined numerical measure of how good the $\pi$ operator is as an
eigenoperator of $t$-$J$ model''.
As mentioned in their paper, the level spacings within each multiplet are
distributed in a relatively small standard deviation for small $J/t$ 
$(0.25,0.5)$,
however, the deviation is very large for large $J/t$ $(1.0,2.0)$ in our
calculations,
which implies that the perturbing correction is more important for large $J/t$.
Concerning the spectral weight of the $\pi$ operator, the intensity of 
the coherent low energy peak decreases with decreasing $J/t$ as in Fig. $1$
and in Ref. \cite{Eder}, which implies that
the perturbing correction is more important for small $J/t$.
Since the the perturbing correction does not behave consistently
as the parameter $J/t$ changes, the role of the perturbing correction is 
not clear.
Even considering the fact that the results of the computer experiments,
that is, the numerical calculations
can be interpreted in certain ways because of the finite-size effect etc.,
one still can claim that the approximate SO($5$) symmetry is not good in this
case.

In conclusions, we have found that the spectra of the $\pi$ and $\Delta$
operators are quite similar and the $\pi$ operator approximation is
not better than the $\Delta$ operator approximation for the two-hole ground
state regardless of the lattice-size.
We also argue that the two-dimensional $t$-$J$ model does not have a good
approximate SO($5$) symmetry.

\vskip 0.5cm

The author wishes to express gratitudes to R. Eder for explaining their
work in detail.
This work was supported by the National Science Council, Rep. of China,
Grant Nos. NSC87-2112-M-011-016.

\center{\bf FIGURE CAPTIONS}
\begin{enumerate}
\noindent {\bf Fig.1.}
Spectra of the $\pi_\pm$ operator for $S=1$, $S_z=\pm 1$, ${\bf P}=(\pi,\pi)$
state (left pannels) and spectra of the $\Delta$ operator for $S=0$,
${\bf P}=(0,0)$ state (right pannels) at half-filling for the $18$-site
lattice.
The final states are in the two-hole sector and the ground state
energy is taken as the chemical potential.
\vspace{1cm}

\noindent{\bf FIG.2.}
$W_{\pi_\pm}$, $W_{\pi_z}$ and $W_\Delta$ as functions of $J/t$ for the
$18$-site lattice.
\vspace{1cm}

\noindent{\bf FIG.3.}
Same as FIG. $2$ but for the $20$-site lattice.

\end{enumerate}

\newpage
\begin{center}
Fig.1 ( Park )
\begin{figure}
\includegraphics{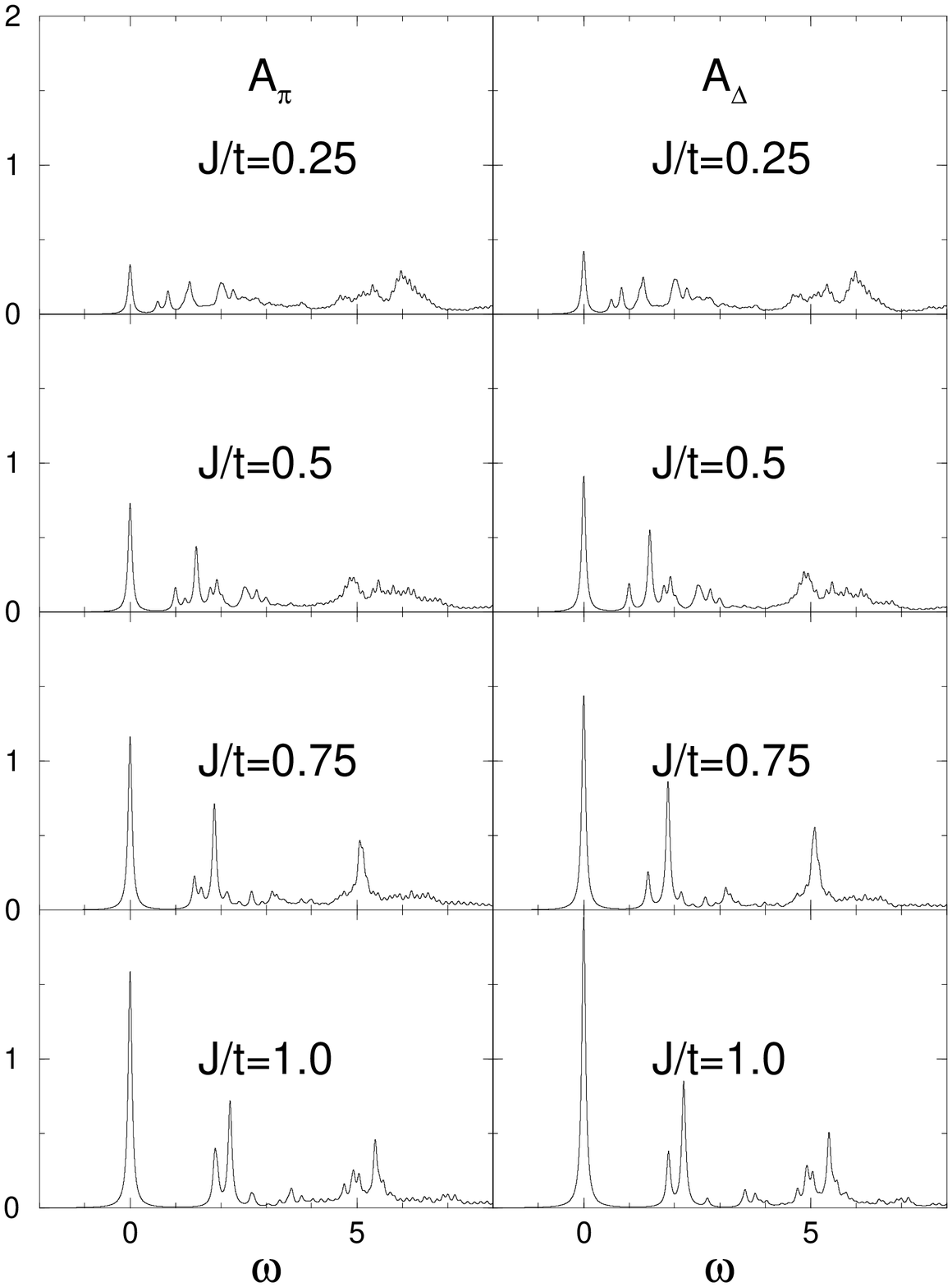}
\end{figure}
\newpage 
Fig.2 ( Park )
\includegraphics{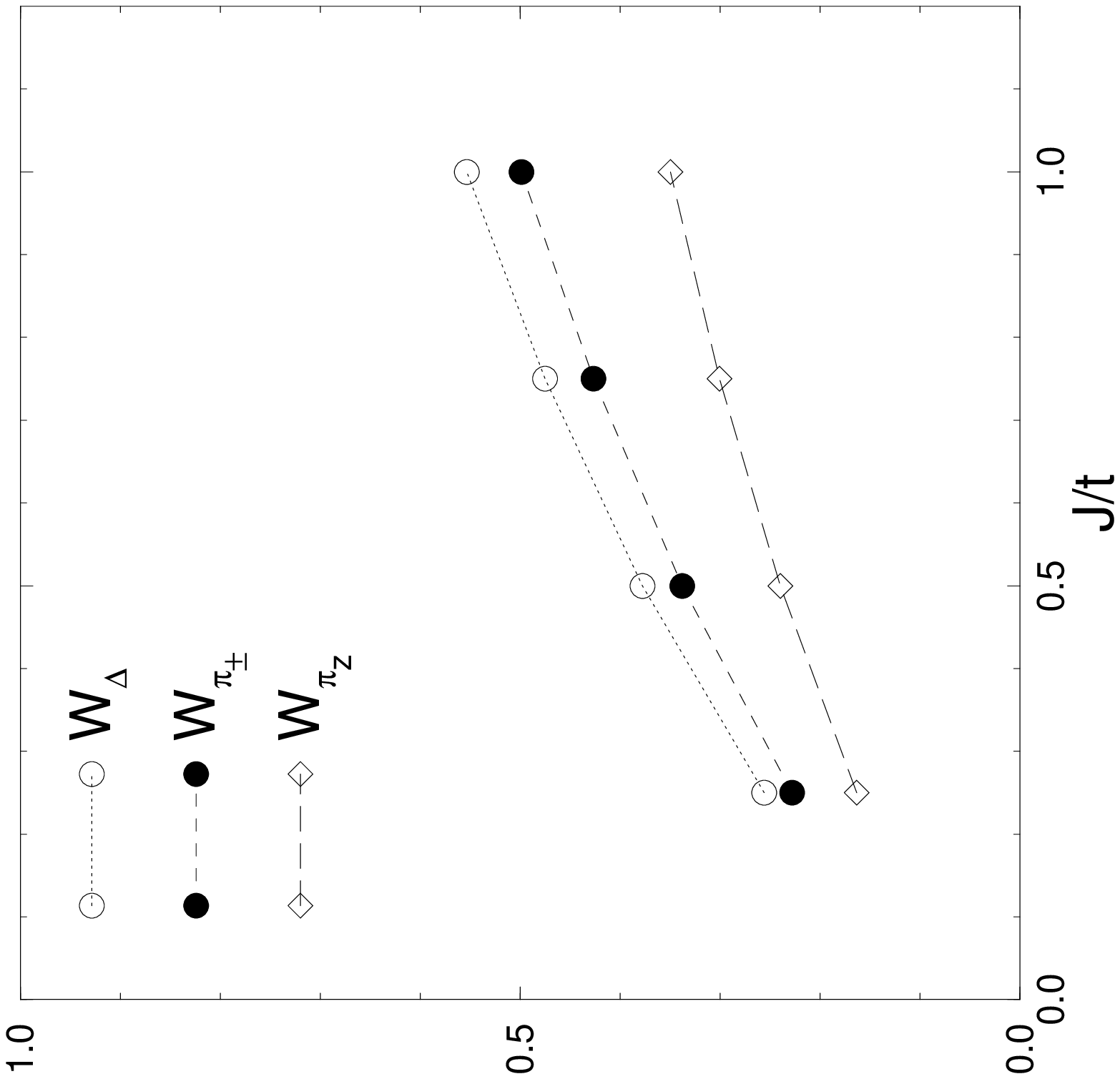}
\newpage 
Fig.3 ( Park )
\includegraphics{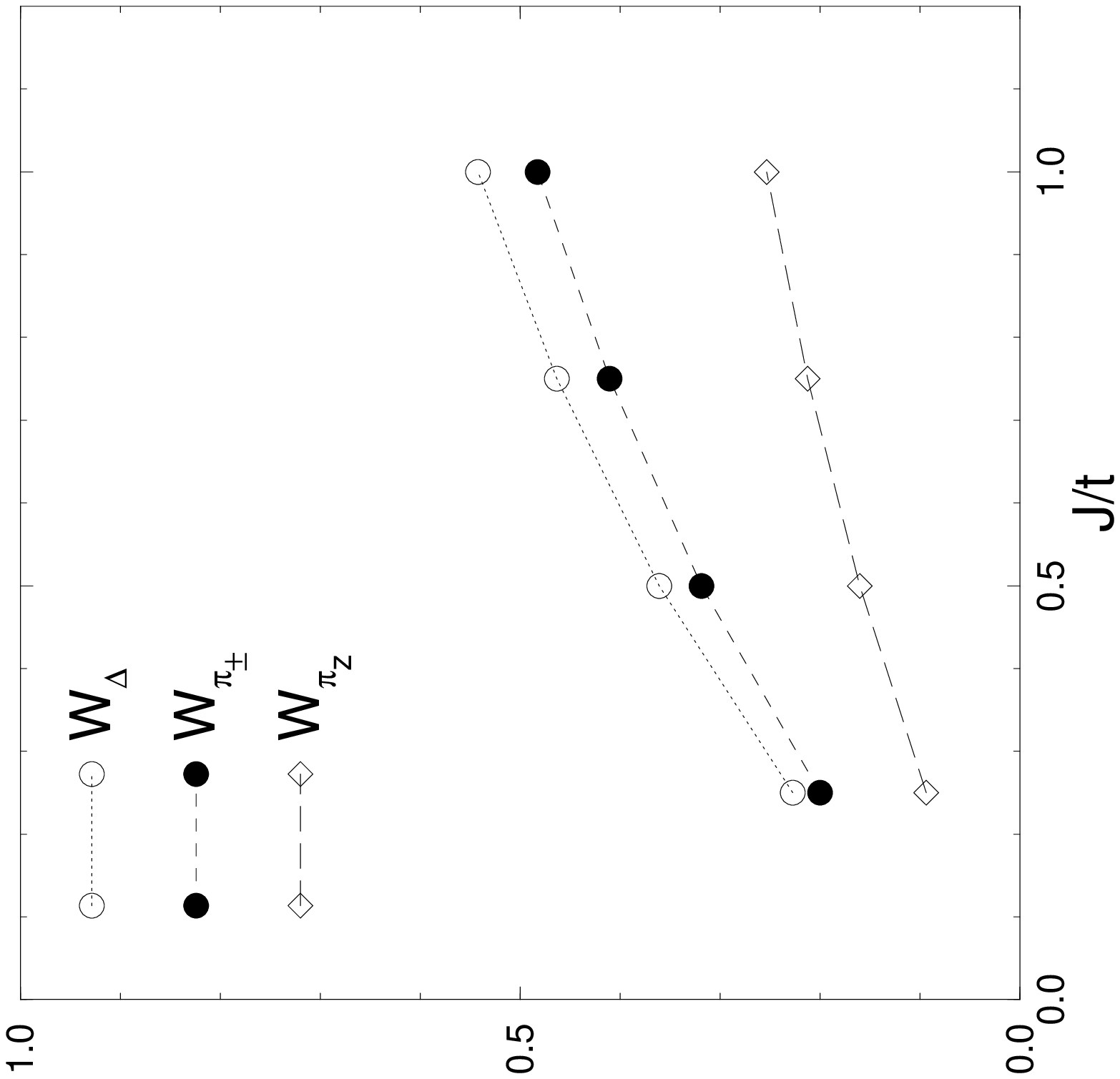}
\end{center}

\end{document}